\PassOptionsToPackage{table}{xcolor}
\documentclass{article}
\usepackage{iclr2027_conference,times}

\usepackage{amsmath,amsfonts,bm}

\def\eqref#1{equation~\ref{#1}}

\def\1{\bm{1}}

\DeclareMathAlphabet{\mathsfit}{\encodingdefault}{\sfdefault}{m}{sl}
\SetMathAlphabet{\mathsfit}{bold}{\encodingdefault}{\sfdefault}{bx}{n}

\usepackage{hyperref}
\usepackage{url}
\hypersetup{hidelinks}
\usepackage{graphicx}
\usepackage{booktabs}
\usepackage{tabularx}
\usepackage{array}
\usepackage{amsmath,amssymb}
\usepackage{xcolor}
\usepackage{microtype}
\usepackage{enumitem}
\usepackage{caption}
\usepackage{float}
\usepackage{cleveref}
\crefname{appendix}{Appendix}{Appendices}
\Crefname{appendix}{Appendix}{Appendices}

\usepackage{multirow,multicol}
\usepackage[normalem]{ulem}
\usepackage{wrapfig}
\usepackage[table]{xcolor}

\newcommand{\vtext}{\mathcal{V}_{\mathrm{text}}}
\newcommand{\vsid}{\mathcal{V}_{\mathrm{SID}}}
\newcommand{\kl}{D_{\mathrm{KL}}}
\newcommand{\base}{\pi_{0}}
\newcommand{\student}{\pi_{\theta}}

\definecolor{bestgreen}{RGB}{212,237,218}
\definecolor{secondblue}{RGB}{230,240,190}
\newcommand{\best}[1]{\cellcolor{bestgreen}\textbf{#1}}
\newcommand{\second}[1]{\cellcolor{secondblue}#1}

\newcommand{\hlbest}[1]{\colorbox{bestgreen}{\textbf{#1}}}
\newcommand{\hlsecond}[1]{\colorbox{secondblue}{#1}}

\title{
Can Generative Retrievers Learn Semantic IDs Without Forgetting How to Speak?}

\author{
Junchen Fu$^{1}$\thanks{This work was completed during an internship at Brave. Correspondence to ron.junchen.fu@gmail.com.} \quad
Kleomenis Katevas$^{2}$ \quad
Vandana Rajan$^{2}$ \quad
Sofía Celi$^{2}$ \quad
Hamed Haddadi$^{2}$ \\
$^{1}$University of Glasgow, Glasgow, United Kingdom \\
$^{2}$Brave, London, United Kingdom \\
}

\iclrfinalcopy
\begin{document}
\maketitle
\begin{abstract}
Generative retrieval (GR) enables end-to-end retrieval by generating document semantic identifiers (SIDs). However, retrieval-only fine-tuning can over-specialize pretrained language models to SID prediction, substantially distorting their natural-language distribution and limiting their suitability for interactive systems that must both retrieve documents and generate natural-language responses. We introduce SpeakGR, a dual-objective framework that learns SIDs while preserving language generation. It combines supervised SID learning with speak-preserving regularization: an on-policy distillation objective that aligns the current model with a frozen copy of the original model on student-generated prefixes using forward KL over the original text vocabulary. We further propose Adaptive SpeakGR, which dynamically adjusts the preservation strength based on observed language drift. Compared with SFT-only, SpeakGR reduces WikiText-2 forward KL by 81.3--93.8\% on MS MARCO and 81.2--85.2\% on Natural Questions (NQ) while retaining effective retrieval across three different LLMs. Adaptive SpeakGR further improves retrieval over SpeakGR in most settings while maintaining substantially
lower language drift than SFT-only.

\end{abstract}

\section{Introduction}

\begin{wrapfigure}{r}{0.45\columnwidth}
    \centering
    \vspace{-8pt}
    \includegraphics[width=\linewidth]{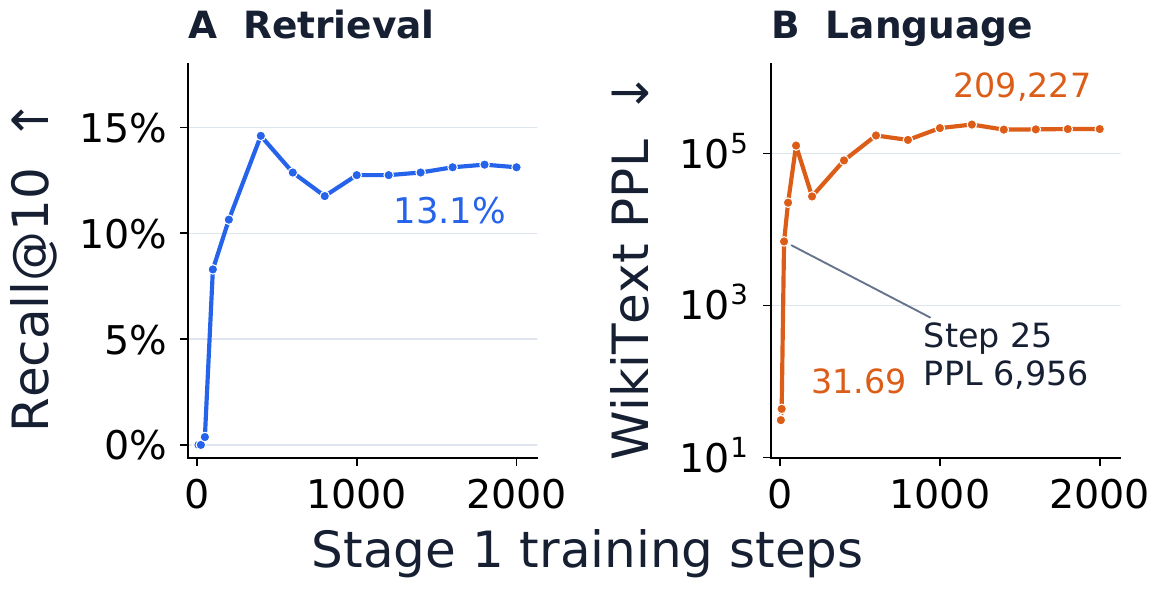}
    \caption{\textbf{Language forgetting during SID indexing.}
Qwen3-0.6B+MS MARCO over 2,000 updates.
(A) Recall@10; (B) WikiText-2 PPL (log scale).
PPL rises from 31.69 to 6,956 within 25 updates. Stage 1 consists of document-to-SID indexing.}
    \label{fig:sft-stage1-forgetting}
    \vspace{-10pt}
\end{wrapfigure}
Generative retrieval (GR), in which a neural model generates document identifiers directly from a query without relying on an explicit search index, has become an active research direction in information retrieval~\citep{tay2022dsi,rajput2023recommender,wang2022nci}. Its central idea is to reformulate retrieval as an \emph{autoregressive generation} problem, replacing conventional candidate scoring with direct prediction of the identifier of a relevant document~\citep{tay2022dsi,wang2022nci,bevilacqua2022seal}. 
A common design identifies each document with a short sequence of semantic
identifiers (SIDs), extends the vocabulary of a pretrained language model
with the SID tokens, and fine-tunes the model to generate these identifiers
autoregressively. This framing turns document retrieval into a native
sequence-generation task and has motivated substantial progress in document
representations, identifier construction, training objectives, and decoding
strategies~\citep{lee2023glen}. However, standard GR objectives optimize the model for SID generation without explicitly preserving its pretrained language-modeling behavior.

Why does this matter? Generative retrieval is attractive when a single language model serves the user end-to-end. The model first retrieves relevant documents by generating their identifiers, then uses the retrieved content to answer, summarize, cite, or continue the interaction~\citep{10.1145/3020165.3020183,10.1145/3331184.3331265,3495724.3496517}. In ambiguous cases, the model may also ask a clarifying question, receive the user's response, and use that clarification to refine retrieval or generate a better answer. This is precisely why retrieval is built on top of a pretrained language model rather than a dedicated encoder. If retrieval fine-tuning substantially degrades the model's language behavior, it can no longer support the broader interaction beyond retrieval, regardless of its Recall@10. Preserving language generation is therefore important for interactive search, where retrieval quality alone does not capture the model's ability to answer follow-up questions, explain retrieved information, or conduct clarification turns with the user.\footnote{In this work, ``language preservation'' specifically refers to preserving the pretrained model's natural-language predictive behavior, rather than guaranteeing preservation of every specialized downstream capability such as mathematical reasoning or code generation.}

Learning SIDs while preserving the pretrained language policy therefore provides three concrete benefits.
\emph{(i) Retrieval as an added capability, not a replacement:} the model learns to emit SIDs while retaining its ability to generate text, allowing retrieval specialization without sacrificing language generation.
\emph{(ii) A single interface for retrieval and for language:} because both SIDs and text are generated autoregressively, a model that keeps both
can support a unified autoregressive interface for retrieval and language generation, with SID resolution connecting retrieval to subsequent generation.
\emph{(iii) Lower system cost:} existing GR systems often
pair the retriever with a separate generator or reader for downstream
knowledge-intensive tasks~\citep{bevilacqua2022seal,song2024re3val}. In contrast, a retriever
that retains strong language capabilities could reuse the same model for
both retrieval and generation, reducing the need to maintain a separate
generation model.

However, none of these benefits and capabilities can be realized, or even studied, until a more basic question is settled: \emph{is the pretrained language behavior retained
during SID-based retrieval specialization?} Standard GR training gives little reason to expect that it does. SID specialization repeatedly trains the same parameters to map documents, pseudo-queries, and queries into short sequences over an identifier vocabulary~\citep{zhou2022ultron,mekonnen2025lightweight, wang2022nci}, with nothing in the objective to constrain the text-token policy. 
Also, most prior work on GR evaluates only retrieval effectiveness~\citep{wang2022nci,tay2022dsi,rajput2023recommender}, leaving the effect on pretrained language behavior largely unknown. In this work, we therefore first measure how SID-based retrieval specialization changes the pretrained language behavior of the model. Across three LLM backbones and two retrieval datasets (six backbone--dataset settings), our experiments reveal a consistent tension: standard SID fine-tuning learns effective retrieval but can rapidly distort the pretrained language distribution, with substantial degradation emerging even at an early stage of training (Figure~\ref{fig:sft-stage1-forgetting}). 
These results further motivate us to ask:
\textbf{Can a generative retriever learn to retrieve with semantic IDs without
forgetting the pretrained language behavior of its backbone?}

To mitigate forgetting of pretrained speaking abilities, we introduce \emph{SpeakGR}, a dual-objective framework combining supervised SID learning with \emph{speak-preserving regularization}. While SID learning specializes the model for GR, the regularizer preserves its pretrained language behavior. Inspired by on-policy distillation (OPD)~\citep{agarwal2024gkd}, the current model generates continuations, while a frozen original model provides next-token distributions at the same prefixes. A forward-KL objective over the original text vocabulary constrains language drift during SID learning. We further introduce \emph{Adaptive SpeakGR}, which adjusts preservation strength according to observed drift.
Across matched model--dataset settings (see~\Cref{tab:all-main}), SpeakGR substantially reduces language drift while retaining effective retrieval. Compared with SpeakGR, Adaptive SpeakGR improves retrieval in five of the six settings and maintains substantially lower language drift than SFT-only.

Our contributions are threefold:
\begin{itemize}
    \item We systematically study forgetting during SID-based generative
    retrieval training and show that standard retrieval specialization can
    substantially distort a pretrained model's natural-language behavior
    despite achieving effective retrieval.

    \item We introduce \emph{SpeakGR}, a dual-objective training framework
    that couples supervised SID learning with \emph{speak-preserving regularization}, allowing retrieval specialization and language preservation to be optimized jointly within the same model. Building on this, we further introduce \emph{Adaptive SpeakGR}, which adjusts the preservation strength in response to the observed language drift.
    
    \item We evaluate SpeakGR and Adaptive SpeakGR across a variety of model--dataset settings and show that they substantially reduce language drift while retaining effective retrieval performance.
\end{itemize}

\section{Related Work}

\paragraph{Generative retrieval.}
Generative retrieval (GR) defines  document retrieval as autoregressive generation
of document identifiers. Representative methods include DSI~\citep{tay2022dsi},
NCI~\citep{wang2022nci}, SEAL~\citep{bevilacqua2022seal}, and
GLEN~\citep{lee2023glen}, which improve identifier construction, generation,
and constrained decoding. More recent work further explores alternative identifier
designs and retrieval objectives, including term-set generation in
TSGen~\citep{zhang2024generative}, simultaneous decoding with hybrid identifiers in
PAG~\citep{zeng2024planning}, and relevance-aware identifier learning in
MERGE~\citep{zhang2025multi}. Related semantic-ID paradigms have also been
adopted in generative recommendation, from learning semantic item identifiers for
autoregressive recommendation~\citep{rajput2023recommender} to recent
differentiable SID learning such as DIGER~\citep{fu2026differentiable}.
While these works primarily focus on improving retrieval or recommendation
effectiveness and identifier learning, we study how SID specialization affects the
pretrained natural-language behavior of the underlying language model.

\paragraph{Preserving language during specialization.}
Fine-tuning on narrow objectives can degrade previously acquired language
capabilities~\citep{11151751}. Existing approaches mitigate such forgetting
through parameter constraints, replay, or distillation
~\citep{doi:10.1073/pnas.1611835114,sun2020distillreplay}. A concurrent work, ORBIT~\citep{verma2026orbit}, studies forgetting in sequential product recommendation, whereas we focus on query-to-document generative retrieval over MS MARCO~\citep{nguyen2016msmarco} and NQ~\citep{kwiatkowski2019natural}. The two methods also differ in preservation strategy: ORBIT monitors parameter drift and triggers origin-model weight merging, while SpeakGR directly regularizes the pretrained language distribution during SID learning. Adaptive SpeakGR further adjusts this preservation strength online according to observed language drift, without weight merging.

More broadly, language preservation during GR specialization remains underexplored, and, to the best of our knowledge, no prior work has explicitly formulated SID learning and language preservation as coupled training objectives. This motivates a unified training framework that learns to retrieve while preserving the model's natural-language behavior.

\section{Problem Formulation}

We consider a pretrained language model that is adapted to generative
retrieval by extending its vocabulary with semantic identifier (SID) tokens.
Let $\base$ denote the frozen pretrained language policy over its original
text vocabulary $\vtext$.

For each retrieval setting, let $L$ denote the fixed identifier length and
let $\mathcal{V}_{\ell}$ denote the SID token set available at position
$\ell\in\{1,\ldots,L\}$. We define the complete SID vocabulary as
\begin{equation}
    \vsid=\bigsqcup_{\ell=1}^{L}\mathcal{V}_{\ell},
    \qquad
    \vsid\cap\vtext=\varnothing,
\end{equation}
where the identifier length and the size of each level-specific vocabulary
are specified by the SID construction used in each experimental setting.

Let $\student$ denote the \emph{current trainable student policy} with
parameters $\theta$. The student is initialized from $\base$, which has its
vocabulary extended from $\vtext$ to $\vtext\cup\vsid$. During retrieval
specialization, $\theta$ is updated while the origin policy $\base$ remains
frozen.

Each indexed document $d$ is assigned an SID using RQ-VAE or another quantization approach based on its text embedding~\citep{tay2022dsi,rajput2023recommender}:
\begin{equation}
    z(d)=(z_1,\ldots,z_L),
    \qquad
    z_{\ell}\in\mathcal{V}_{\ell}.
\end{equation}
For a query $q$, the generative retriever models the probability of an
identifier autoregressively as
\begin{equation}
    \student(z\mid q)
    =
    \prod_{\ell=1}^{L}
    \student
    \left(
        z_{\ell}
        \mid
        q,z_{<\ell};
        \mathcal{V}_{\ell}
    \right),
    \label{eq:sidfactorization}
\end{equation}
where
$\student(\cdot\mid q,z_{<\ell};\mathcal{V}_{\ell})$
denotes the student's next-token distribution after restricting and
renormalizing the logits to the valid SID tokens in
$\mathcal{V}_{\ell}$. Thus, although the student is defined over the enlarged
vocabulary $\vtext\cup\vsid$, retrieval operates over a level-specific SID
action space.

\subsection{Retrieval specialization}

Let $\mathcal{Y}$ denote the set of supervised query--identifier pairs
$(q,z)$. If a query is associated with multiple relevant documents, each
query--document association contributes a separate pair with the
corresponding document identifier.

Following~\cite{wang2022nci}, we optimize a level-restricted cross-entropy objective over the supervised
SID positions:
\begin{equation}
    \mathcal{L}_{\mathrm{ret}}(\theta)
    =
    -\frac{1}{N_{\mathrm{ret}}}
    \sum_{(q,z)\in\mathcal{Y}}
    \sum_{\ell=1}^{L}
    \log
    \student
    \left(
        z_{\ell}
        \mid
        q,z_{<\ell};
        \mathcal{V}_{\ell}
    \right),
    \qquad
    N_{\mathrm{ret}}
    =
    \sum_{(q,z)\in\mathcal{Y}} |z|
    =
    L|\mathcal{Y}|.
    \label{eq:retloss}
\end{equation}

The loss is normalized by the number of supervised SID tokens.
Each training sequence has the form
$(q,z_1,\ldots,z_L,\texttt{EOS})$.
The \texttt{EOS} token is teacher-forced for sequence-format consistency but
is excluded from~\Cref{eq:retloss}, since the identifier length is
fixed within each retrieval setting.

At inference time, SID generation is constrained by the level-specific
action spaces together with a trie constructed from the identifiers of the
indexed documents. The trie ensures that decoding produces a valid and
complete document identifier. Each valid decoded identifier maps to a document bucket;
documents sharing an SID are ranked within that bucket.

\subsection{Measuring Language Drift}
\label{sec:language-drift}

Retrieval specialization updates a model that was originally pretrained to
predict natural-language tokens. To quantify the resulting language forgetting (a form of catastrophic forgetting), we
measure how far the specialized model's next-token distribution deviates from
that of the frozen pretrained model on natural text.

For any natural-language prefix $s$, let
$\student^{\vtext}(\cdot\mid s)$ denote the student's next-token distribution
restricted to the original text vocabulary $\vtext$ and renormalized:
\begin{equation}
    \student^{\vtext}(v\mid s)
    =
    \frac{\student(v\mid s)}
    {\sum_{u\in\vtext}\student(u\mid s)},
    \qquad
    v\in\vtext.
    \label{eq:textpolicy}
\end{equation}
This restriction isolates changes within the original text-token distribution
from probability mass assigned directly to the newly introduced SID tokens.

Let $\mu_{\mathrm{text}}$ denote a distribution over natural-language
prefixes. We quantify language drift as follows:

\begin{equation}
    \Delta_{\mathrm{lang}}(\student,\base)
    =
    \mathbb{E}_{s\sim\mu_{\mathrm{text}}}
    \left[
        \kl\!\left(
            \base(\cdot\mid s)
            \,\middle\|\,
            \student^{\vtext}(\cdot\mid s)
        \right)
    \right],
    \label{eq:langdef}
\end{equation}

where KL is the Kullback--Leibler divergence.
A smaller $\Delta_{\mathrm{lang}}$ indicates that the specialized model
remains closer to the pretrained model's next-token behavior on natural text.
In our experiments, we instantiate $\mu_{\mathrm{text}}$ with fixed
prefixes from the WikiText-2 test split~\citep{merity2016pointer}
and use $\Delta_{\mathrm{lang}}$ as our primary measure of
distributional preservation.

\begin{figure}[t]
    \centering
    \includegraphics[width=0.75\linewidth]{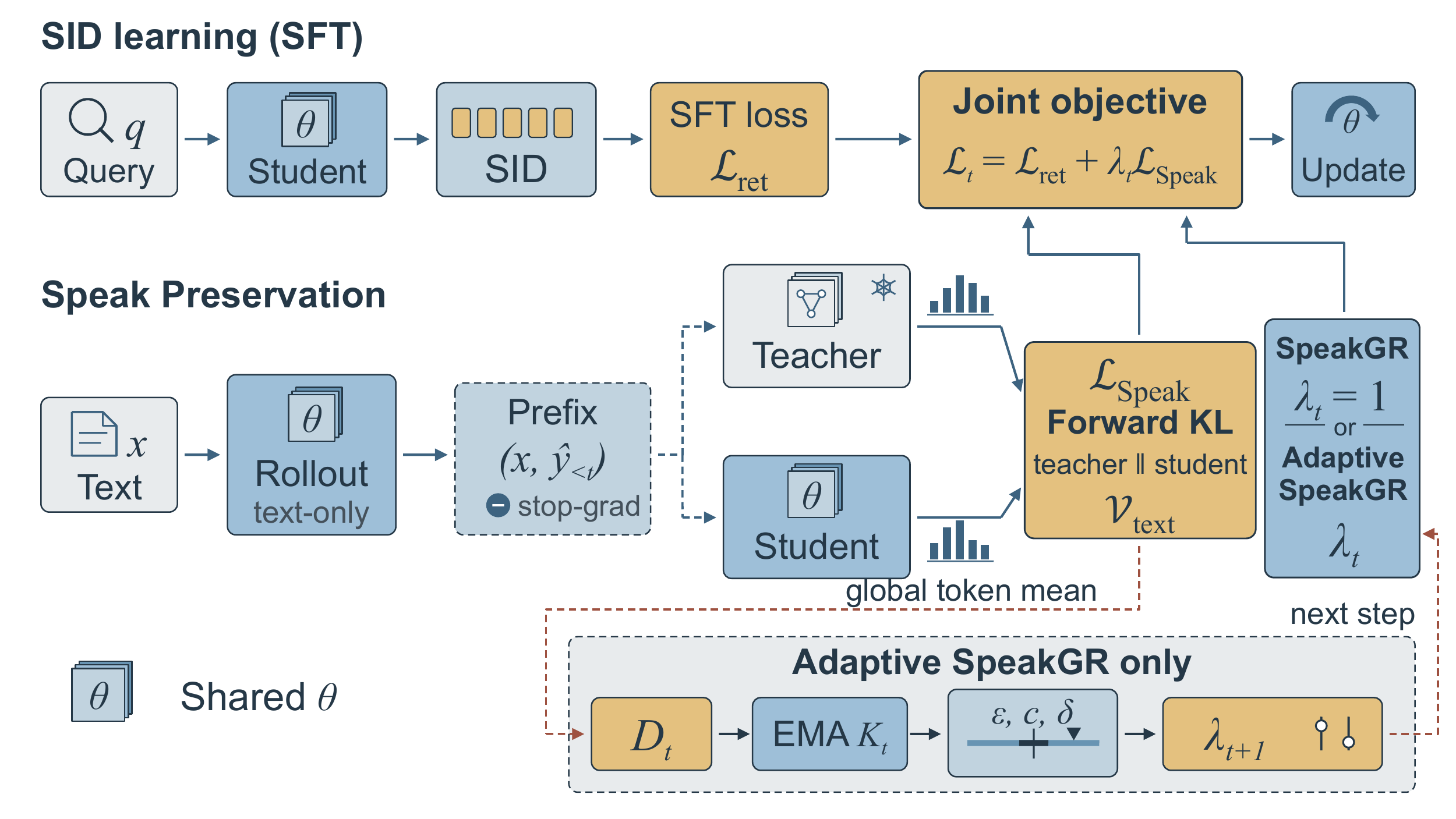}
    \vspace{-0.1in}
    \caption{\textbf{Overview of SpeakGR and its adaptive variant.}
A shared student learns SID retrieval while preserving language behavior
through forward KL from a frozen teacher at student-sampled text prefixes.
SpeakGR fixes the preservation weight at $\lambda_t=1$; Adaptive SpeakGR uses the dashed controller to update $\lambda_{t+1}$ from observed
preservation KL for the next step.}
    \label{fig:overview}
    \vspace{-0.2in}
\end{figure}
\section{SpeakGR: Joint SID Learning and speak-preserving regularization}
\label{sec:speakgr}

\paragraph{One model, two objectives.}
The goal of \emph{SpeakGR} is to jointly optimize two objectives over the same student parameters
$\theta$: (i) an SID-learning objective for effective generative retrieval and (ii) a
\emph{speak-preserving regularization} objective used to retain the model's pretrained
natural-language behavior. The decoding space is specified by
the application: retrieval restricts generation to the level-specific SID
vocabulary and the document trie, whereas natural-language generation uses
the original text vocabulary $\vtext$. This, in short, means that both objectives train the same
model parameters for their respective behaviors.

\subsection{Speak-preserving regularization}
\label{subsec:regul}

Inspired by on-policy distillation~\citep{agarwal2024gkd}, we regularize the current model on natural-language states that it visits during generation.
Let $\mathcal{D}_{\mathrm{lang}}$ denote the fixed pool of
natural-language prompts used for preservation, and let
$\widetilde{\pi}_{\theta}$ denote the rollout distribution obtained from the
current student after masking all SID logits and sampling only from $\vtext$.

For a prompt $x\in\mathcal{D}_{\mathrm{lang}}$, the current student samples a text-only continuation:
\begin{equation}
    \hat y
    \sim
    \widetilde{\pi}_{\theta}(\cdot\mid x; \vtext).
\end{equation}

The sampled tokens are treated as constants, so no gradient passes through
the discrete rollout. The frozen pretrained model and the trainable student then receive the same concatenated sequence $(x,\hat y)$, each producing a next-token
distribution at every response position.

For a preservation minibatch
$\mathcal{B}_{\mathrm{lang}}\subset\mathcal{D}_{\mathrm{lang}}$, we match the
two models at every sampled response position using forward KL over the
original text vocabulary (where $T_i$ is the length of the continuation the student sampled for prompt $i$):
\begin{equation}
\begin{split}
    \mathcal{L}_{\mathrm{Speak}}(\theta)
    =
    \frac{1}{N_{\mathrm{lang}}}
    \sum_{i\in\mathcal{B}_{\mathrm{lang}}}
    \sum_{t=1}^{T_i}
    \sum_{v\in\vtext}
    \base(v\mid s_{i,t})
    \log
    \frac{\base(v\mid s_{i,t})}
         {\student^{\vtext}(v\mid s_{i,t})},
    \\[-2pt]
    s_{i,t}=(x_i,\hat y_{i,<t}),
    \qquad
    N_{\mathrm{lang}}
    =
    \sum_{i\in\mathcal{B}_{\mathrm{lang}}}T_i.
    \label{eq:speak}
\end{split}
\end{equation}

The current student determines the prefixes $s_{i,t}$ on which preservation
is performed. The frozen model does not generate target sequences; instead,
it provides its full next-token distribution at the same model-generated
prefixes. This allows \emph{speak-preserving regularization} to regularize the model on the
natural-language states it encounters as SID specialization progresses.

We use forward KL because, at a fixed prefix, minimizing
$\kl[\base\|\student^{\vtext}]$ is equivalent, up to the entropy of the
frozen model, to minimizing cross-entropy under its token distribution.
Forward KL therefore penalizes the student for assigning insufficient
probability to tokens favored by the pretrained model. We use forward KL to preserve the pretrained text distribution; we do not imply that it is universally preferable to other divergence measures.

$\mathcal{L}_{\mathrm{Speak}}$ constrains the student's natural language next-token
distribution, but does not directly optimize any specific downstream task capability. As we will see in~\Cref{sec:results}, it reverses most of the distributional damage caused by SID specialization. 

\subsection{SpeakGR Training}

SpeakGR combines supervised SID learning with \emph{speak-preserving regularization} (see~\Cref{subsec:regul}) through a
dual-objective training loss:
\begin{equation}
    \mathcal{L}_{\mathrm{SpeakGR}}
    =
    \mathcal{L}_{\mathrm{ret}}
    +
    \mathcal{L}_{\mathrm{Speak}}.
    \label{eq:speakgr}
\end{equation}
The two losses are separately token-normalized. The retrieval objective drives
specialization toward SID generation, while the preservation objective
regularizes natural-language behavior during the same optimization process.

\subsubsection{Adaptive SpeakGR}
\label{sec:adaptive-speakgr}

A static preservation weight applies the same pressure throughout training and cannot adapt to changes in language drift. An intuitive
alternative is to strengthen preservation when the model drifts further from
its pretrained behavior and relax it when the drift is small. Adaptive
SpeakGR follows this principle as it dynamically adjusts only the
preservation coefficient $\lambda_t$, while leaving the rest of SpeakGR
unchanged.

At optimizer step $t$, let
$\mathcal{B}_{\mathrm{lang},t}$ denote the current preservation minibatch and
define
\begin{equation}
    D_t
    :=
    \mathcal{L}_{\mathrm{Speak}}
    (\theta_t;\mathcal{B}_{\mathrm{lang},t})
\end{equation}
as the globally token-averaged preservation KL observed at that step.
We maintain its exponential moving average (EMA)~\citep{brown1959statistical},
and update the coefficient multiplicatively:

\begin{align}
    K_t
    &=\alpha K_{t-1}+(1-\alpha)D_t,\\
    e_t
    &=\operatorname{clip}
      \left(K_t/\epsilon-1,-c,c\right),\\
    \lambda_{t+1}
    &=\operatorname{clip}_{[\lambda_{\min},\lambda_{\max}]}
      \left(
      \lambda_t\exp(\eta\,\tilde e_t)
      \right),
    \label{eq:controller}
\end{align}
where
\begin{equation}
    \tilde e_t =
    \begin{cases}
        0, & |e_t|\leq\delta,\\
        e_t, & \text{otherwise}.
    \end{cases}
\end{equation}

Here, $\alpha\in[0,1)$ is the EMA smoothing factor, $\epsilon>0$ is the
reference preservation-KL level, $c$ bounds the normalized deviation
$e_t$, and $\delta$ defines a deadband around zero. The adaptation rate
$\eta$ controls how quickly $\lambda_t$ changes, while
$\lambda_{\min}$ and $\lambda_{\max}$ bound its allowable range. The current optimizer step uses $\lambda_t$, while the observed preservation
KL determines the coefficient for the next step. The resulting objective is
\begin{equation}
    \mathcal{L}_{t}
    =
    \mathcal{L}_{\mathrm{ret}}
    +
    \lambda_t\mathcal{L}_{\mathrm{Speak}}
    \label{eq:adaptive-speakgr}
\end{equation}

Adaptive SpeakGR therefore changes only the strength of
speak-preserving regularization over training, leaving the underlying dual-objective
formulation unchanged. Details of the controller hyperparameters are given
in Appendix~\ref{sec:exp-setup}.

\section{Experiments}
\label{sec:experiments}

Our experiments aim to answer three questions:

\begin{enumerate}
    \item How effectively can SpeakGR mitigate the language drift caused by SID specialization while retaining effective SID retrieval?

    \item How does SpeakGR compare with other language-preservation strategies for generative retrieval?

    \item How does adaptive preservation affect the retrieval--preservation trade-off, and how sensitive is it to the reference level $\epsilon$, which directly sets the controller's target preservation-KL level?
\end{enumerate}

We compare retrieval-only SFT, SpeakGR, and Adaptive SpeakGR
across two retrieval corpora and three backbones, measuring both retrieval
quality and language drift.
We evaluate offline replay and ORBIT-style weight
merging under a matched protocol.
Throughout, retrieval and language metrics are computed on fixed held-out
sets that play no role in training or checkpoint selection. We first describe the datasets and evaluation protocol, then present the results in~\Cref{sec:results}. Additional experimental details are provided in~\Cref{sec:exp-setup}, and the evaluation metrics in~\Cref{sec:eval-metrics}. We further discuss our results in~\Cref{sec:discussion}.

\providecommand{\best}[1]{\textcolor{green!45!black}{#1}}
\providecommand{\second}[1]{\textcolor{blue!75!black}{#1}}
\begin{table*}[t]
\caption{Retrieval and language-preservation results across three backbones on MS MARCO and NQ. The best and second-best results among adapted models within each dataset--backbone setting are highlighted in \hlbest{green} and \hlsecond{light green}, respectively. AdaSpeakGR denotes Adaptive SpeakGR.}
\vspace{-0.12in}
\label{tab:all-main}
\centering\footnotesize
\renewcommand{\arraystretch}{0.9}
\setlength{\tabcolsep}{4pt}
\resizebox{\textwidth}{!}{\begin{tabular}{@{}cll|rrr|rrrr@{}}
\toprule
Dataset & Backbone & Method & R@1 $\uparrow$ & R@10 $\uparrow$ & MRR@10 $\uparrow$ & FKL $\downarrow$ & PPL $\downarrow$ & Top-1 $\uparrow$ & SID mass $\downarrow$ \\
\midrule
\multirow{12}{*}{\textbf{MS MARCO}} & Qwen3-0.6B & Base & -- & -- & -- & 0.000 & 31.69 & 100.00\% & $\approx0$ \\
 &  & SFT-only & \second{0.2426} & \second{0.6386} & \second{0.3649} & 8.103 & 53,966.98 & 0.13\% & 4.16\% \\
 &  & SpeakGR & 0.2364 & 0.6349 & 0.3571 & \best{1.519} & \best{88.48} & \best{46.15\%} & \second{0.62\%} \\
 &  & AdaSpeakGR & \best{0.2537} & \best{0.6498} & \best{0.3725} & \second{1.728} & \second{105.86} & \second{43.24\%} & \best{0.55\%} \\
\cmidrule(l){2-10}
 & Qwen3-1.7B & Base & -- & -- & -- & 0.000 & 23.07 & 100.00\% & $\approx0$ \\
 &  & SFT-only & \best{0.2760} & \second{0.6696} & \best{0.3977} & 11.222 & 648,108.29 & 0.54\% & 4.18\% \\
 &  & SpeakGR & 0.2562 & 0.6324 & 0.3764 & \best{1.761} & \best{66.72} & \best{48.05\%} & \best{0.15\%} \\
 &  & AdaSpeakGR & \second{0.2649} & \best{0.6745} & \second{0.3927} & \second{1.841} & \second{70.66} & \second{46.88\%} & \second{0.17\%} \\
\cmidrule(l){2-10}
 & Gemma-3-1B-IT & Base & -- & -- & -- & 0.000 & 102.38 & 100.00\% & $\approx0$ \\
 &  & SFT-only & 0.2525 & 0.6337 & 0.3808 & 13.005 & 13,615,975.95 & 1.20\% & 26.37\% \\
 &  & SpeakGR & \second{0.2624} & \second{0.6361} & \second{0.3813} & \second{0.809} & \second{88.15} & \second{56.40\%} & \second{0.20\%} \\
 &  & AdaSpeakGR & \best{0.2686} & \best{0.6399} & \best{0.3864} & \best{0.775} & \best{86.10} & \best{57.87\%} & \best{0.19\%} \\
\midrule
\multirow{12}{*}{\textbf{NQ}} & Qwen3-0.6B & Base & -- & -- & -- & 0.000 & 31.69 & 100.00\% & $\approx0$ \\
 &  & SFT-only & \best{0.4713} & \best{0.7397} & \best{0.5618} & 6.379 & 10,251.46 & 6.07\% & 7.00\% \\
 &  & SpeakGR & 0.3960 & 0.6582 & 0.4786 & \best{1.044} & \best{56.30} & \best{52.40\%} & \second{1.91\%} \\
 &  & AdaSpeakGR & \second{0.4706} & \second{0.7285} & \second{0.5543} & \second{1.116} & \second{66.51} & \second{52.33\%} & \best{1.51\%} \\
\cmidrule(l){2-10}
 & Qwen3-1.7B & Base & -- & -- & -- & 0.000 & 23.07 & 100.00\% & $\approx0$ \\
 &  & SFT-only & \best{0.4808} & \best{0.7507} & \best{0.5701} & 5.545 & 2,452.24 & 16.53\% & 8.60\% \\
 &  & SpeakGR & 0.4056 & 0.7172 & 0.5084 & \second{1.042} & \second{35.74} & \best{58.11\%} & \second{0.23\%} \\
 &  & AdaSpeakGR & \second{0.4387} & \second{0.7280} & \second{0.5324} & \best{1.030} & \best{34.66} & \second{57.84\%} & \best{0.20\%} \\
\cmidrule(l){2-10}
 & Gemma-3-1B-IT & Base & -- & -- & -- & 0.000 & 102.38 & 100.00\% & $\approx0$ \\
 &  & SFT-only & 0.3925 & 0.7138 & \second{0.4961} & 5.631 & 7,791.61 & 6.70\% & 18.32\% \\
 &  & SpeakGR & \best{0.4126} & \best{0.7324} & \best{0.5169} & \second{0.831} & \second{91.58} & \second{55.96\%} & \second{0.34\%} \\
 &  & AdaSpeakGR & \second{0.3932} & \second{0.7189} & 0.4956 & \best{0.743} & \best{86.04} & \best{58.04\%} & \best{0.30\%} \\
\bottomrule
\end{tabular}}
\vspace{-0.25in}
\end{table*}

\subsection{Datasets}

\paragraph{Retrieval datasets and training.}
Following \citet{mekonnen2025lightweight} and \citet{sun2023learning},
we use retrieval subsets derived from MS MARCO \citep{nguyen2016msmarco}
and NQ \citep{kwiatkowski2019natural}. Our processed corpora contain
319,872 and 109,739 documents, respectively; we refer to them as
MS MARCO and NQ throughout.
Following \cite{zhou2022ultron} and \cite{mekonnen2025lightweight}, training proceeds in three stages: document-to-SID indexing, synthetic query-to-SID training, and supervised query-to-SID fine-tuning (see \Cref{tab:sftconfig} for detailed configurations). These stages learn the document
catalog, broaden query coverage, and adapt to human queries.
For synthetic queries, we use ten released docT5query generations
per document on MS MARCO \citep{nogueira2019doc2query}
and the released synthetic queries on NQ~\citep{sun2023learning}.

\paragraph{Language-preservation prompts.}
Our pool contains 2,048 continuation prefixes from WikiText-2
training text \citep{merity2016pointer} and 2,048 instruction-style
prompts derived from training passages, MS MARCO's training questions,
and task templates. Motivated by the importance of instruction diversity
\citep{wang2023self}, this mixture provides varied contexts
for preserving the base model's language behavior, covering both
free-text continuation and instruction-following.
SpeakGR and Adaptive SpeakGR use only prompts:
the student generates continuations, and the frozen base model
provides next-token distributions on the same prefixes.
WikiText-2 test is reserved exclusively for evaluation.
Pool composition, construction details, and examples appear in~\cref{app:language-prompt-pool}
(\Cref{tab:language-prompt-composition}
and~\Cref{tab:language-prompt-examples}).

\subsection{Results}
\label{sec:results}

\subsubsection{Retrieval and language preservation}

To address the first question, we compare retrieval-only SFT with SpeakGR and Adaptive
SpeakGR across six matched dataset--backbone settings: MS MARCO and NQ
with Qwen3-0.6B, Qwen3-1.7B, and Gemma-3-1B-IT. For each setting, all
methods start from the same pretrained model, use the same SID assignments
and retrieval supervision, and are evaluated with the same retrieval and
WikiText-2 protocols. \Cref{tab:all-main} reports both retrieval metrics
and measures of language preservation.

Retrieval-only SFT consistently causes substantial language drift. Across the
six settings, its WikiText forward KL~(FKL) ranges from $5.545$ to $13.005$.
SpeakGR reduces this to $0.809$--$1.761$, corresponding to relative reductions
of $81.3$--$93.8\%$ on MS MARCO and $81.2$--$85.2\%$ on NQ. The same trend
is reflected in perplexity~(PPL) and top-1 agreement. Retrieval effects depend on the
setting: SpeakGR remains close to SFT on MS MARCO with Qwen3-0.6B, loses more recall
on the Qwen NQ settings, and improves recall for both Gemma settings. Adaptive
SpeakGR often recovers additional retrieval performance: for example, on MS
MARCO with Qwen3-1.7B it reaches an R@10 of $0.6745$ versus $0.6696$ for SFT-only while
reducing FKL from $11.222$ to $1.841$.

\paragraph{What this tells us.}
SpeakGR consistently preserves the pretrained natural-language distribution
far better than retrieval-only SFT while retaining effective SID retrieval.
The exact retrieval--preservation balance varies by dataset and backbone, and
the adaptive variant can recover additional retrieval performance in several
settings without giving up the preservation benefit.\footnote{Achieving state-of-the-art retrieval performance is beyond the scope of this paper; stronger retrieval methods are complementary and could be combined with speak-preserving regularization~\citep{sun2023learning,mekonnen2025lightweight}. 
} Further analysis shows that the degradation is driven primarily by changes in the model’s relative probabilities over ordinary text tokens, rather than by probability mass leaking from text tokens to the newly introduced SID vocabulary (\Cref{sec:mechanism}). Representation analysis further shows that speak-preserving regularization substantially reduces directional drift in middle and later layers while still allowing internal representations to adapt for retrieval.

\begin{table*}[t]
\caption{Comparison with language-preservation baselines on Qwen3-0.6B
across MS MARCO and NQ. In ORBIT, the fraction of original parameters whose signs differ from the pretrained model is used to trigger back-merging when the threshold $\tau$ is exceeded. LP, MP, and HP denote low, medium, and high preservation constraint strength, respectively; smaller $\tau$ imposes a stronger constraint. The best and second-best results among adapted models within each dataset are highlighted in \hlbest{green} and \hlsecond{light green}, respectively. AdaSpeakGR denotes Adaptive SpeakGR.}
\vspace{-0.12in}
\label{tab:preservation-baselines}
\centering
\footnotesize
\setlength{\tabcolsep}{5pt}
\renewcommand{\arraystretch}{0.95}
\resizebox{\textwidth}{!}{\begin{tabular}{cl | rrr|rrrr}
\toprule
Dataset & Method
& R@1 $\uparrow$ & R@10 $\uparrow$ & MRR@10 $\uparrow$
& FKL $\downarrow$ & PPL $\downarrow$ & Top-1 $\uparrow$
& SID mass $\downarrow$ \\
\midrule

\multirow{8}{*}{\textbf{MS MARCO}}
& Base
& -- & -- & -- & 0.000 & 31.69 & 100.00\% & $\approx 0$ \\

& SFT-only
& 0.2426 & 0.6386 & 0.3649
& 8.103 & 53,966.98 & 0.13\% & 4.16\% \\

& SpeakGR
& 0.2364 & 0.6349 & 0.3571
& \second{1.519} & \second{88.48}
& \second{46.15\%} & 0.62\% \\

& AdaSpeakGR
& 0.2537 & \second{0.6498} & \second{0.3725}
& 1.728 & 105.86 & 43.24\% & 0.55\% \\

& Offline Replay
& 0.2475 & 0.5903 & 0.3531
& 8.672 & 32,933.41 & 28.58\% & \second{0.116\%} \\

& ORBIT (LP, $\tau=0.10$)
& \best{0.2946} & \best{0.6584} & \best{0.4119}
& 7.391 & 35,443.46 & 3.57\% & 7.576\% \\

& ORBIT (MP, $\tau=0.05$)
& \second{0.2611} & 0.5817 & 0.3657
& 3.927 & 1,262.86 & 17.55\% & 0.931\% \\

& ORBIT (HP, $\tau=0.007$)
& 0.0309 & 0.1980 & 0.0731
& \best{0.273} & \best{42.84}
& \best{76.33\%} & \best{0.013\%} \\

\midrule

\multirow{8}{*}{\textbf{NQ}}
& Base
& -- & -- & -- & 0.000 & 31.69 & 100.00\% & $\approx 0$ \\

& SFT-only
& \best{0.4713} & \best{0.7397} & \best{0.5618} & 6.379 & 10,251.46 & 6.07\% & 7.00\% \\

& SpeakGR
& 0.3960 & 0.6582 & 0.4786 & \second{1.044} & \second{56.30} & \second{52.40\%} & 1.91\% \\

& AdaSpeakGR
& \second{0.4706} & \second{0.7285} & \second{0.5543} & 1.116 & 66.51 & 52.33\% & \second{1.51\%} \\

& Offline Replay
& 0.3075 & 0.5811 & 0.3906
& 4.445 & 2,420.93 & 26.06\% & 5.681\% \\

& ORBIT (LP, $\tau=0.10$)
& 0.4406 & 0.7236 & 0.5327
& 6.257 & 9,208.21 & 6.73\% & 8.337\% \\

& ORBIT (MP, $\tau=0.05$)
& 0.3799 & 0.6656 & 0.4713
& 4.110 & 1,251.65 & 21.36\% & 6.570\% \\

& ORBIT (HP, $\tau=0.007$)
& 0.1775 & 0.3849 & 0.2394
& \best{0.500} & \best{43.73}
& \best{67.52\%} & \best{0.272\%} \\
\bottomrule
\end{tabular}}
\vspace{-0.2in}
\end{table*}

\subsubsection{Comparison with language-preservation baselines}

To address the second question, we compare SpeakGR with two representative alternatives for
retaining previous behavior during specialization: offline replay and
ORBIT. Replay adds a response-level language CE objective from a fixed
buffer, whereas ORBIT monitors parameter drift and conditionally merges the
student weights with the frozen origin model. We evaluate all methods on
Qwen3-0.6B for both NQ and MS MARCO in~\Cref{tab:preservation-baselines}.

The baselines expose different retrieval--preservation trade-offs. At the evaluated setting, offline replay provides limited protection of the held-out language distribution: on
MS MARCO, it reaches an FKL of $8.672$ and an R@10 of $0.5903$. ORBIT is strongly threshold-dependent: a loose threshold retains retrieval performance but leaves substantial language drift (R@10/FKL $0.6584/7.391$ on MS MARCO), whereas a tight threshold
reduces FKL to $0.273$ but also reduces R@10 to $0.1980$. SpeakGR occupies a
different region of this trade-off, with Adaptive SpeakGR retaining an R@10 of $0.6498$
and an FKL of $1.728$ on MS MARCO. The same qualitative pattern appears on NQ: Adaptive SpeakGR
achieves an R@10 of $0.7285$ with an FKL of $1.116$, compared with $0.5811/4.445$ for replay and $0.7236/6.257$ for
ORBIT with a loose threshold, reported as R@10/FKL.

\paragraph{What this tells us.}
No preservation method dominates across all operating points. At its
evaluated setting, replay reduces language drift relative to retrieval-only
SFT on NQ but does not match SpeakGR's joint retrieval--preservation balance.
ORBIT can move between retrieval and preservation through its threshold,  
but incurs a non-trivial retrieval cost under strong correction. SpeakGR instead provides a direct objective-level alternative that
achieves substantial language preservation while maintaining strong retrieval performance,
with Adaptive SpeakGR offering a stronger retrieval-oriented operating point.

\subsection{Ablation and adaptive sensitivity}

\begin{wrapfigure}{r}{0.45\textwidth}
    \centering
    \vspace{-8pt}
    \includegraphics[width=\linewidth]{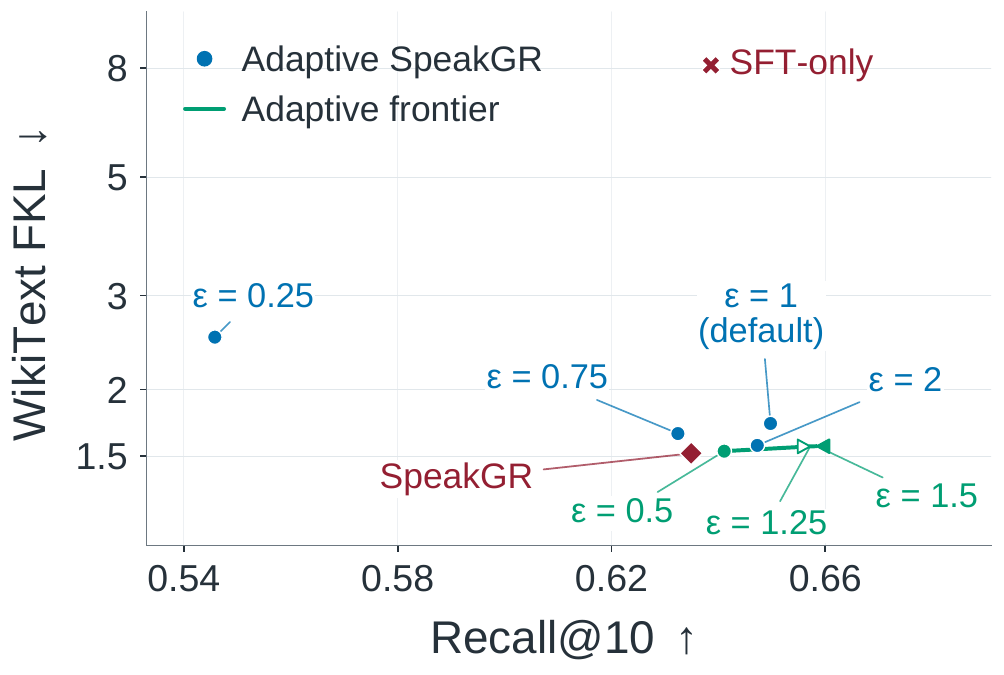}
    \vspace{-1pt}
    \caption{\textbf{Retrieval--preservation trade-off on Qwen3-0.6B+MS MARCO.}
Lower right is better. Labels indicate $\epsilon$; the star marks
the default $\epsilon=1$, and the diamond denotes SpeakGR.
The green line is the Pareto frontier, which connects Adaptive SpeakGR runs not outperformed
by another $\epsilon$ in both objectives.
FKL uses a log scale; triangle tips locate near-coincident points.}
    \label{fig:pareto}
    \vspace{-15pt}
\end{wrapfigure}

To address the third question, we vary the adaptive controller’s reference preservation level $\epsilon\in \{0.25,0.50,0.75,1.00,1.25,1.50,2.00\}$ on Qwen3-0.6B with MS MARCO.
Every point is a three-stage run and uses the same checkpoint-selection
rule based on a separate Stage-3 validation set.
\Cref{tab:ablation_kl_budget} reports the numerical results, while~\Cref{fig:pareto} shows the retrieval--preservation trade-off.

The response to $\epsilon$ is non-monotonic. The strictest setting
($\epsilon=0.25$) hurts both retrieval and held-out preservation, reaching
R@10 $0.5458$ and FKL $2.507$. In contrast, moderate values between $0.5$ and
$2.0$ produce a substantially better region of the trade-off; the best R@10 is $0.6584$ at $\epsilon=1.5$ with FKL $1.566$.  Importantly,
a smaller training reference level does not necessarily yield a smaller
held-out WikiText FKL because the controller observes student-generated
training states rather than the held-out evaluation distribution. We additionally study sensitivity to a fixed preservation
coefficient in~\Cref{app:fixed_weight}. The results
show that SpeakGR remains effective across a broad range of
fixed weights, while Adaptive SpeakGR removes the need to
commit to a single fixed coefficient by adjusting its
preservation strength online.

\paragraph{What this tells us.}
Adaptive SpeakGR does not require a narrowly tuned reference level in a
moderate range, but an overly strict target can be counterproductive. The
adaptive mechanism therefore provides a practical way to move along the
retrieval--preservation trade-off, while $\epsilon$ should be interpreted as a
training reference rather than a direct bound on held-out language drift.

\section{Conclusion}
We ask a simple question: can a language model learn to retrieve with semantic IDs without forgetting how to generate natural language? Our experiments show that retrieval-only fine-tuning can substantially distort the pretrained language distribution, but this degradation can be substantially mitigated. SpeakGR jointly optimizes SID learning and speak-preserving regularization, reducing WikiText forward-KL drift while retaining effective retrieval performance, with trade-offs that vary across datasets and backbones. 
Adaptive SpeakGR adjusts preservation strength in response to the observed drift, offering an alternative to fixed weighting. Together, our results show that SID-based generative retrieval can be learned while substantially limiting drift from the model’s pretrained natural-language distribution.

\section*{Reproducibility Statement}

Code is available at the anonymous repository linked in the abstract. The experimental section and appendices provide the training schedule, model and optimization settings, Adaptive SpeakGR hyperparameters, preservation-prompt construction, and retrieval and language evaluation protocols needed to reproduce the reported experiments.

\section*{AI Assistance}

The core research idea was developed by the authors. Generative AI tools assisted with identifying relevant literature, generating code for experimental implementation, curating part of the language-preservation prompt pool, assisting figures preparation, and drafting and polishing manuscript text. The authors reviewed the AI-assisted materials and take full responsibility for the methods, results, and final manuscript.

\subsubsection*{Acknowledgments}
We thank Alexandros Karatzoglou and Joemon M. Jose for their valuable feedback throughout this project.

\bibliography{references}
\bibliographystyle{plainnat}
\clearpage

\appendix
\crefalias{section}{appendix}

\begin{table}[t]
\centering
\small
\setlength{\tabcolsep}{12pt}
\caption{Sensitivity of Adaptive SpeakGR to the reference preservation level $\epsilon$ on Qwen3-0.6B with MS MARCO. The $\epsilon=1$ result is reused from the main table as the default setup.}
\label{tab:ablation_kl_budget}
\begin{tabular}{crr}
\toprule
Reference KL $\epsilon$
& R@10 $\uparrow$
& Wiki FKL $\downarrow$ \\
\midrule
0.25 & 0.5458 & 2.507 \\
0.50 & 0.6411 & 1.533 \\
0.75 & 0.6324 & 1.654 \\
1.00 & 0.6498 & 1.728 \\
1.25 & 0.6572 & 1.565 \\
1.50 & 0.6584 & 1.566 \\
2.00 & 0.6473 & 1.572 \\
\bottomrule
\end{tabular}
\end{table}

\section{Experimental Setup}
\label{sec:exp-setup}

\begin{table}[ht]
\caption{The three-stage retrieval schedule. Effective global batches and data
order are held fixed across GPU counts through gradient accumulation.}
\label{tab:sftconfig}
\centering
\small
\begin{tabular}{c| l rrrr}
\toprule
Stage & Mapping & Steps & LR & Global batch & Max length \\
\midrule
1 & document $\rightarrow$ SID & 2,000 & $2\times10^{-4}$ & 1,536 & 320 \\
2 & pseudo-query $\rightarrow$ SID & 8,000 & $5\times10^{-4}$ & 3,072 & 144 \\
3 & query $\rightarrow$ SID & 2,000 & $1\times10^{-4}$ & 1,536 & 144 \\
\bottomrule
\end{tabular}
\end{table}

\subsection{Models and training}

We use Qwen3-0.6B/1.7B \citep{yang2025qwen3} and Gemma-3-1B-IT
\citep{team2025gemma}, fully fine-tuning all student parameters,
including embeddings, while keeping the base teacher frozen.
We tune the number and size of the RQ-VAE codebooks, selecting five
256-entry codebooks to quantize Qwen3-Embedding document representations
\citep{zhang2025qwen3,rajput2023recommender}.
SIDs contain five tokens.
Within each dataset--backbone comparison, all methods use identical
SID assignments and student initialization. Only 1.94\% of MS MARCO SID buckets contain multiple documents;
we use a frozen Qwen3-Embedding-0.6B encoder to rerank documents
within colliding SID buckets on both datasets while preserving
the generated bucket order.

\Cref{tab:sftconfig} summarizes the primary retrieval schedule,
with the longest stage devoted to pseudo-query coverage.
We use AdamW \citep{loshchilov2017decoupled}, 3\% linear warmup followed
by linear decay, and six language prompts per optimizer update. We generate student continuations with SID tokens masked and a temperature of 0.8,
top-$p=0.9$, top-$k=20$, and at most 32 tokens. SpeakGR minimizes $\mathcal{L}_{\mathrm{ret}}+\mathcal{L}_{\mathrm{Speak}}$.
In all main settings, Adaptive SpeakGR uses
$\epsilon=1$ nat/token, $\alpha=0.95$, $c=0.2$,
$\delta=0.05$, $\eta=0.05$, and $\lambda_{\max}=1$.
Stage~1 starts with $\lambda_0=0.1$ and
$\lambda_{\min}=0.01$.
Qwen3-0.6B retains this lower bound and carries
the controller state across stages.
For Qwen3-1.7B and Gemma-3-1B-IT, Stage~2 resets the
controller to $\lambda=0.25$ with a fresh KL EMA;
Stages~2--3 use $\lambda_{\min}=0.25$, and Stage~3
inherits the Stage~2 controller state.
The controller observes training-stream KL.

\paragraph{Baselines.}
Following prior work on replay for continual language learning
\citep{sun2020distillreplay}, we use an offline language-replay baseline.
Its numerical settings are chosen to match SpeakGR's language-data and
sequence-length budget: replay uses six
examples per optimizer update, retains at most 640 prompt tokens, and
supervises at most 32 response tokens per example. The replay buffer uses the same prompt sources and fixed offline
responses described in Appendix~\Cref{app:language-prompt-pool}. Its response-only cross-entropy
is computed over the original text vocabulary, excluding added SID
tokens from the softmax normalization and masking prompt and padding
positions. To match SpeakGR's auxiliary-loss weighting, the response-only replay loss is
added to the retrieval loss with coefficient
$\lambda_{\mathrm{replay}}=1.0$, the same coefficient used for SpeakGR's
language-preservation loss. We fix this
configuration as a matched reference and do not claim that it is optimal. 
We reimplement ORBIT following its original paper
\citep{verma2026orbit}, because no official implementation was publicly
available at the time of our experiments. Following the original algorithm,
whenever sign dissimilarity exceeds its threshold, the student weights are
averaged $1{:}1$ with the corresponding base-model weights, excluding SID
rows and leaving the optimizer states unchanged. In~\Cref{tab:preservation-baselines}, we denote the ORBIT threshold by $\tau$ to distinguish it from Adaptive SpeakGR's reference level $\epsilon$.

\subsection{Evaluation}

We report Recall@1/10 and MRR@10 using trie-constrained beams of 10
for MS MARCO and 100 for NQ following~\cite{mekonnen2025lightweight} and \cite{sun2023learning}.
Language evaluation scores 16,384 next-token predictions in 64
non-overlapping, 256-token WikiText-2 test input blocks, with shifted
targets and reset context. Every causal position is scored.
We measure fidelity to the base model using forward KL and
top-1 agreement \citep{stanton2021does}, and additionally report
perplexity. These metrics are computed after renormalizing over
the base-model vocabulary. We also report SID probability mass
before masking. Comparisons are made within each
dataset--backbone setting.

\section{Discussion}
\label{sec:discussion}

\paragraph{Why WikiText?}
We use WikiText-2 as the primary language-preservation evaluation because it provides a fixed held-out distribution of natural text on which the specialized model can be compared directly with its pretrained counterpart token by token. This gives a controlled measure of distributional drift, but should not be interpreted as a complete measure of language capability. We use this as our primary measure of distributional preservation; broader capability preservation is outside the scope of our evaluation.

\paragraph{Can one model both retrieve and generate?}
Preserving natural-language behavior is particularly relevant when the same model is expected to support both retrieval and generation. In such a setting, the model can first generate an SID under retrieval constraints, resolve it to a document, and then use the retrieved content to produce a natural-language response. Our current experiments establish preservation of the pretrained text distribution during retrieval specialization, but do not yet constitute a full end-to-end retrieve--resolve--generate evaluation. We therefore view this unified use case as a motivating application rather than a demonstrated capability.

\paragraph{Hyperparameter sensitivity.}
Our study does not exhaustively tune the regularization strength, rollout sampling parameters, replay settings, or the adaptive controller. Accordingly, the replay result represents one fixed operating point rather than a tuned replay frontier. SpeakGR uses a simple static preservation weight, while
Adaptive SpeakGR adjusts this weight online according to the observed preservation
KL. Although this design provides a useful adaptive mechanism, improvements
cannot be attributed solely to adaptation without comparing against a static
SpeakGR run using a matched average preservation weight. We therefore treat
the adaptive variant as an additional operating strategy rather than claiming
that adaptive weighting is universally superior.

\paragraph{Limitations.}
Our main evaluation covers three language-model backbones and two retrieval
corpora, but most cross-setting comparisons are based on a single main seed
and one SID construction family. Moreover, strong recovery of the WikiText-2 distribution should not be interpreted as complete preservation of broader capabilities. Our evaluation is intentionally scoped to pretrained natural-language distribution preservation during retrieval specialization. Specialized capabilities such as mathematical reasoning, coding, and task-specific problem solving are outside the scope of this study. The displayed SpeakGR anchors in~\Cref{tab:preservation-baselines} match those in~\Cref{tab:all-main}. Finally, broader
multi-seed evaluation and stronger matched controls for the adaptive
coefficient remain necessary before making claims of universal capability
preservation or Pareto superiority.

\section{Mechanism analysis}
\label{sec:mechanism}
\paragraph{Preserving the language distribution.}
For a natural-language prefix distribution $\mu$, the preservation penalty is
\begin{equation}
    R_\mu(\theta)=\mathbb{E}_{s\sim\mu}
    \kl\!\left[\base(\cdot\mid s)\,\|\,\student^{\vtext}(\cdot\mid s)\right].
    \label{eq:speak-function-regularizer}
\end{equation}
At a fixed prefix, its gradient with respect to the student text logits is
$\student^{\vtext}(\cdot\mid s)-\base(\cdot\mid s)$, correcting probability
differences from the frozen base model. Define the shared predictive mass as
$M(s)=\sum_{v\in\vtext}\min\{\base(v\mid s),\student^{\vtext}(v\mid s)\}$.
Since $M(s)=1-\mathrm{TV}(\base,\student^{\vtext})$, Pinsker's
inequality~\citep{canonne2022short} and Jensen's inequality yield
\begin{equation}
    \mathbb{E}_{s\sim\mu}M(s)
    \geq\max\!\left\{0,\,1-\sqrt{R_\mu(\theta)/2}\right\}.
    \label{eq:speak-overlap-bound}
\end{equation}
Reducing KL therefore raises a lower bound on shared probability mass
on the same prefix distribution; transfer to held-out text is measured
empirically.

\begin{figure}[t]
    \centering
    \includegraphics[width=0.8\linewidth]{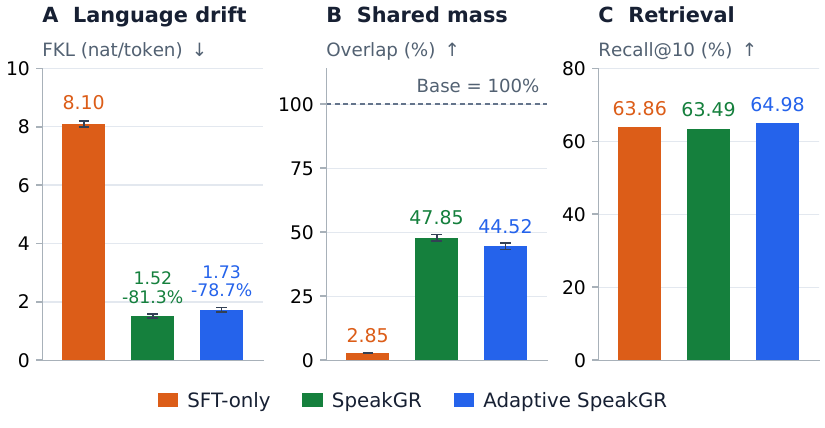}
    \caption{\textbf{Language preservation with SpeakGR and Adaptive SpeakGR.}
    Selected Qwen3-0.6B checkpoints on MS MARCO.
    (A) Forward KL from the frozen base model to the student; percentages indicate reductions
    relative to SFT-only. (B) Shared predictive probability mass with the base
    model. (C) Recall@10 on 808 retrieval queries. A/B use the same
    $64\times256$ WikiText-2 test positions; 95\% confidence intervals use
    10,000 block-bootstrap resamples conditional on the selected checkpoints.}
    
    \label{fig:speakgr-mechanism}
\end{figure}

On the WikiText evaluation in Figure~\ref{fig:speakgr-mechanism}, SpeakGR and
Adaptive SpeakGR reduce FKL from SFT's $8.10$ to $1.52$ and $1.73$, while
increasing shared mass from $2.85\%$ to $47.85\%$ and $44.52\%$.
SID leakage accounts for only $0.58\%$ of SFT's excess full-vocabulary NLL;
the main degradation is within the normalized text distribution.

\begin{figure}[t]
    \centering
    \includegraphics[width=\linewidth]{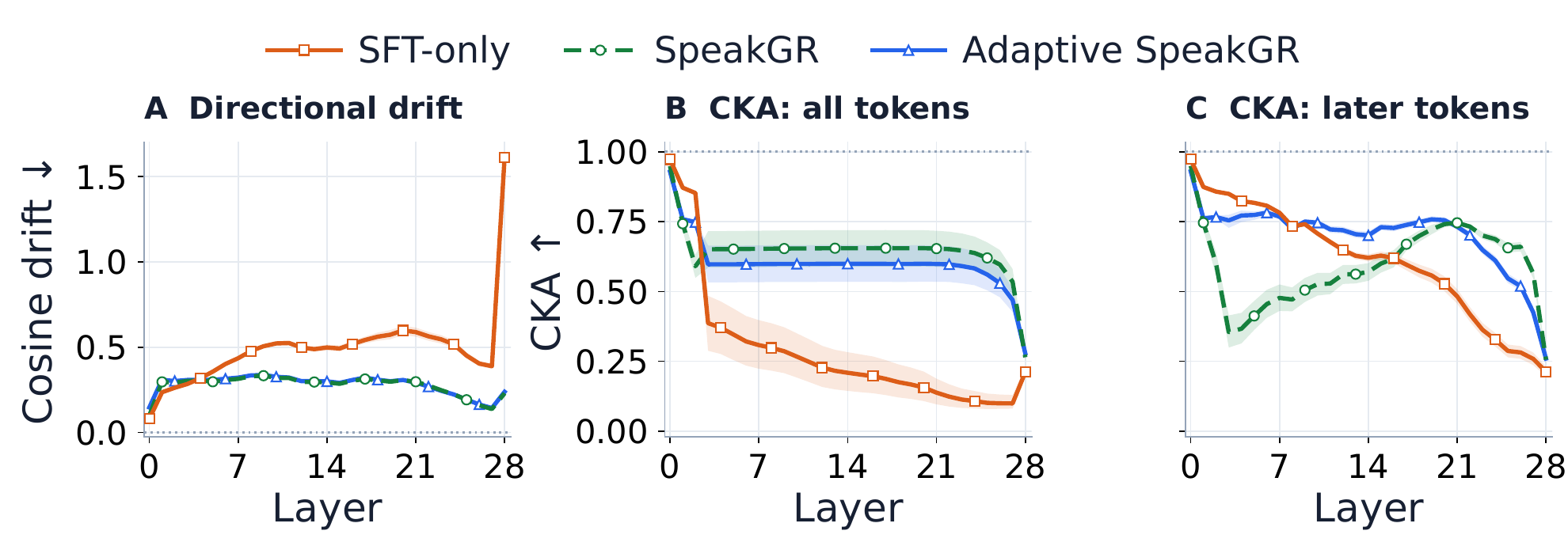}
    \caption{\textbf{Layer-wise residual-stream comparison on
    Qwen3-0.6B+MS MARCO.}
    (A) Cosine drift from the base model; (B) centered linear CKA;
    (C) CKA excluding the first 16 positions per block, with context unchanged.
    Layer 0 is the embedding output; layers 1--28 are block outputs before
    the final RMS normalization. A/B use the same 16,384 WikiText positions;
    C uses 15,360 of these positions.
    Bands show pointwise 95\% block-bootstrap intervals.}
    \label{fig:speakgr-layerwise}
\end{figure}

\paragraph{Where do the representations change?}
On the same checkpoints and prefixes, we compare base and student residuals
$h_0^{(j)}(s)$ and $h_\theta^{(j)}(s)$ using mean cosine drift
$1-\cos(h_0^{(j)}(s),h_\theta^{(j)}(s))$ and centered linear
CKA~\citep{kornblith2019similarity}. CKA compares representation geometry
up to orthogonal rotations and uniform rescaling, computed within each
text block and then averaged.

\Cref{fig:speakgr-layerwise}A shows lower directional drift through
the middle and later layers with preservation. At the final block, drift
is $1.61$ for SFT, $0.23$ for SpeakGR, and $0.24$ for Adaptive SpeakGR;
these values are nearly unchanged when the first 16 positions of each block
are excluded. The geometry comparison is more sensitive to position selection
(B/C): at layer 14, SpeakGR versus SFT CKA changes from $0.65$ versus $0.21$
to $0.57$ versus $0.62$ after this exclusion. Early layers are also not
uniformly closer to the base. Thus, the evidence supports directional
alignment in later layers, rather than a claim that all representation
geometry is preserved.

These observations are consistent with function-space regularization:
preserving language predictions permits internal adaptation for retrieval.
This endpoint comparison does not identify a causal layer of forgetting
or establish complete downstream capability preservation.

\section{Language-preservation prompt pool}
\label{app:language-prompt-pool}

The complete pool combines natural-text continuation prefixes with
instruction-style prompts. This mixture provides contexts for preserving
both ordinary text generation and responses to instructions and questions.
In~\Cref{tab:language-prompt-composition}, we show the full
composition, and in~\Cref{tab:language-prompt-examples}, we illustrate the inputs.
These categories describe training prompts, not independently evaluated capabilities. They are included to diversify natural-language contexts for preservation.

\begin{table}[!htbp]
\centering
\caption{\textbf{Composition of the complete 4,096-entry preservation-prompt
pool.} Counts refer to stored prompt instances, not unique templates.
Source groups are shown in italics.}
\label{tab:language-prompt-composition}
\small
\setlength{\tabcolsep}{5pt}
\renewcommand{\arraystretch}{1.12}
\begin{tabularx}{\linewidth}{@{}>{\raggedright\arraybackslash}Xr@{}}
\toprule
Prompt category & Count \\
\midrule
\multicolumn{2}{@{}l}{\emph{A. Continuation prefixes: WikiText-2 training split}} \\
Natural-text continuation & 2,048 \\
\midrule
\multicolumn{2}{@{}l}{\textbf{B. Instruction-style prompts}} \\
\multicolumn{2}{@{}l}{\emph{B.1. WikiText-2 training text + task instructions (1,208)}} \\
Summarization & 218 \\
Explanation & 218 \\
Paraphrasing & 212 \\
Key-point extraction & 218 \\
Title generation & 217 \\
Translation & 125 \\
\addlinespace[3pt]
\multicolumn{2}{@{}l}{\emph{B.2. MS MARCO training-query QA (217)}} \\
Direct QA & 111 \\
Document-conditioned QA & 106 \\
\addlinespace[3pt]
\multicolumn{2}{@{}l}{\emph{B.3. Programmatically instantiated task templates (623)}} \\
Writing (email) & 120 \\
Arithmetic & 184 \\
Logic & 97 \\
Coding & 110 \\
Advice/list generation & 112 \\
\midrule
\textbf{Instruction subtotal} & \textbf{2,048} \\
\textbf{Complete pool} & \textbf{4,096} \\
\bottomrule
\end{tabularx}

\vspace{4pt}
\begin{minipage}{\linewidth}
\footnotesize
\end{minipage}
\end{table}

\paragraph{Construction and supervision.}
Passage-based prompts attach fixed instructions to WikiText-2 training
text. QA prompts use MS MARCO training questions, either alone or with
retrieved evidence. Other prompts instantiate task templates with varying
inputs, such as numbers for arithmetic and entities for logic.
Candidates are screened using offline responses and are selected by task
category; arithmetic and logic use rule-based reference responses during
preparation. SpeakGR and Adaptive SpeakGR consume only the selected prompts,
not their stored answers: the student generates continuations and the frozen
base model scores the same prefixes.

\begin{table}[!htbp]
\centering
\caption{\textbf{Representative examples from the preservation-prompt
pool.} Examples are abbreviated from actual prompts and lightly formatted
for readability. Bracketed fields replace passages or evidence; ellipses
mark omitted continuation text. Continuation inputs are raw text prefixes,
not explicit instructions to continue writing.}
\label{tab:language-prompt-examples}
\small
\setlength{\tabcolsep}{5pt}
\renewcommand{\arraystretch}{1.15}
\begin{tabularx}{\linewidth}{@{}>{\raggedright\arraybackslash}p{0.23\linewidth}>{\raggedright\arraybackslash}X@{}}
\toprule
Category & Example prompt (abbreviated) \\
\midrule
Continuation &
Virginia Tech running back Cyrus Lawrence finished the game\ldots \\
\addlinespace[3pt]
Summarization &
Summarize the passage in two concise sentences.
Passage: \emph{[WikiText-2 training passage]} \\
\addlinespace[3pt]
Translation &
Translate the following English sentence into Chinese.
English: The church's north and south walls are supported by a series
of buttresses. \\
\midrule
Direct QA &
Question: what is the arctic desert.
Answer the question in one or two clear sentences. \\
\addlinespace[3pt]
Document-conditioned QA &
Question: define statistical inference.
Evidence: \emph{[retrieved document]}.
Use only the provided evidence. \\
\midrule
Writing &
Write exactly three short sentences to Sam, politely asking to reschedule
a meeting because of a seminar. \\
\addlinespace[3pt]
Arithmetic &
A store has 80 items and packs them equally into 4 boxes.
How many full items go in each box and how many remain? \\
\addlinespace[3pt]
Logic &
Premise: All salmon are fish. Some fish are red.
Can we conclude that some salmon are red? Explain in two sentences. \\
\addlinespace[3pt]
Coding &
Write a Python function that counts word frequencies in a string.
Return only the code, without explanation. \\
\addlinespace[3pt]
Advice/list generation &
List three practical ways of saving energy at home.
Keep each item concise. \\
\bottomrule
\end{tabularx}
\end{table}

\section{Evaluation Metrics}
\label{sec:eval-metrics}

\paragraph{Retrieval.}
Recall@1 and Recall@10 measure the fraction of queries whose
relevant document appears in the top-1 and top-10 results. Each evaluation query is scored against a single designated relevant document.
MRR@10 averages the reciprocal rank of the relevant document,
assigning zero when it falls outside the top 10.
Higher values indicate better retrieval.

\paragraph{Language preservation.}
All language metrics are averaged over the same WikiText-2
evaluation positions. Using $\base$ and the text-normalized
student $\student^{\vtext}$ from \Cref{eq:textpolicy},
we report:
\textbf{FKL}, the base-to-student forward KL defined in
\Cref{eq:langdef};
\textbf{PPL}, the exponential of the student's mean
negative log-likelihood of the observed next tokens; and
\textbf{Top-1 agreement}, the percentage of positions where
the two models predict the same highest-probability token.
Lower FKL/PPL and higher agreement are better.

\paragraph{SID mass.}
We report the average percentage of probability assigned to
SID tokens, $100\sum_{v\in\vsid}\student(v\mid s)$,
before text-vocabulary masking. Lower values indicate less
probability allocated to identifiers in language contexts.
Language metrics are compared within each dataset--backbone setting.

\section{Fixed-Weight Sensitivity}
\label{app:fixed_weight}

\Cref{tab:fixed_lambda_sensitivity} examines whether
SpeakGR depends on a carefully chosen static preservation
coefficient. Across $\lambda \in \{0.1, 0.5, 1.0, 1.5\}$,
all non-zero settings substantially reduce WikiText FKL,
from 8.103 for retrieval-only SFT to 1.435--1.532, while
retaining effective retrieval performance. This indicates
that the preservation effect is robust to the choice of a
fixed coefficient rather than depending on a single
carefully tuned value. Different coefficients nevertheless
lead to different retrieval--preservation operating points,
showing that the choice of preservation strength still
affects the balance between the two objectives.

Notably, appropriately chosen fixed coefficients achieve
some of the strongest results in this sensitivity study.
We therefore do not claim that adaptive weighting is
intrinsically superior to a well-chosen fixed weight.
Instead, Adaptive SpeakGR is motivated by a practical
difference in how the preservation strength is controlled.
Static SpeakGR requires a single coefficient to be selected
before training and uses the same preservation coefficient throughout training. Adaptive SpeakGR instead adjusts $\lambda_t$ online according to the observed preservation KL, allowing the regularization pressure to change as language drift evolves. Its intended advantage is therefore not a guaranteed improvement over the best fixed operating point, but reduced reliance on committing to one manually selected preservation coefficient for the entire training process.

\paragraph{What this tells us.} Together with the sensitivity analysis in~\Cref{tab:ablation_kl_budget}, these results show that
speak-preserving regularization is effective and robust under
fixed weighting, while Adaptive SpeakGR provides a
self-adjusting alternative for controlling the
retrieval--preservation trade-off. When a suitable fixed
coefficient is known, static weighting can be a strong
choice; when the appropriate preservation strength is less
clear or may vary over training, the adaptive formulation
offers a practical mechanism for adjusting that strength
online in response to the observed language drift.

\begin{table}[h]
\centering
\small
\caption{Sensitivity to the fixed SpeakGR preservation coefficient on
Qwen3-0.6B with MS MARCO. Best results are highlighted in \hlbest{green}.}
\label{tab:fixed_lambda_sensitivity}
\resizebox{\linewidth}{!}{\begin{tabular}{lrrrrrrr}
\toprule
$\lambda$
& R@1 $\uparrow$
& R@10 $\uparrow$
& MRR@10 $\uparrow$
& FKL $\downarrow$
& PPL $\downarrow$
& Top-1 $\uparrow$
& SID mass $\downarrow$ \\
\midrule
$0$ (SFT-only)
& 0.2426 & 0.6386 & 0.3649
& 8.103 & 53,966.98 & 0.13\% & 4.16\% \\

$0.1$
& 0.2512 & \best{0.6621} & 0.3804
& \best{1.435} & 82.02 & \best{48.11\%} & 0.66\% \\

$0.5$
& \best{0.2723} & 0.6535 & \best{0.3904}
& 1.474 & \best{80.21} & 47.31\% & \best{0.54\%} \\

$1.0$ (SpeakGR)
& 0.2364 & 0.6349 & 0.3571
& 1.519 & 88.48 & 46.15\% & 0.62\% \\

$1.5$
& 0.2364 & 0.6262 & 0.3493
& 1.532 & 89.31 & 46.55\% & 0.63\% \\
\bottomrule
\end{tabular}}
\vspace{2pt}
\end{table}

\end{document}